\documentclass[conference]{IEEEtran}
\IEEEoverridecommandlockouts
\usepackage{cite}
\usepackage{amsmath,amssymb,amsfonts}
\usepackage{algorithmic}
\usepackage{graphicx}
\usepackage{textcomp}
\usepackage{xcolor}
\usepackage{amsfonts}
\usepackage{booktabs}
\usepackage{float}
\usepackage{graphicx}
\usepackage{grffile}
\usepackage{longtable}
\usepackage{wrapfig}
\usepackage{gensymb}
\usepackage{textcomp}
\usepackage{amsmath}
\usepackage{algorithm}
\usepackage{verbatim}
\usepackage{graphicx}
\usepackage{caption}
\usepackage{listings}
\usepackage{array}
\usepackage{subcaption}
\usepackage{booktabs}
\usepackage{amsmath}
\usepackage{enumitem}
\usepackage{dblfloatfix}
\usepackage{balance}
\usepackage{lastpage}
\usepackage[utf8]{inputenc}

\title{Intelligent Network Layer for Cyber-Physical Systems Security}

\author{\IEEEauthorblockN{Raj Chaganti\IEEEauthorrefmark{1}, Deepti Gupta\IEEEauthorrefmark{2}, Naga Vemprala\IEEEauthorrefmark{3}}
\IEEEauthorblockA{\IEEEauthorrefmark{1}CSIRT, ExpediaGroup Inc\\\IEEEauthorrefmark{2}Dept. of Computer Science,
University of Texas at San Antonio,
San Antonio, Texas 78249, USA \\\IEEEauthorrefmark{3}
University of Texas at San Antonio,
San Antonio, Texas 78249, USA}
\IEEEauthorrefmark{1}raj.chaganti2@gmail.com, 
\IEEEauthorrefmark{2}deepti.mrt@gmail.com,
\IEEEauthorrefmark{3}naga.vemprala@gmail.com}

\begin{document}

\maketitle
\begin{abstract}
Cyber-Physical System (CPS) has made a tremendous progress in recent years and also disrupted many technical fields such as smart industries, smart health, smart transportation etc. to flourish the nations economy. However, CPS Security is still one of the concerns for wide adoption owing to high number of devices connecting to the internet and the traditional security solutions may not be suitable to protect the advanced, application specific attacks. This paper presents a programmable device network layer architecture to combat attacks and efficient network monitoring in heterogeneous environment CPS applications. We leverage Industrial control systems (ICS) to discuss the existing issues, highlighting the importance of advanced network layer for CPS. The programmable data plane language (P4) is introduced to detect well known HELLO Flood attack with minimal efforts in the network level and also used to featuring the potential solutions for security.
\end{abstract}


\section{Introduction}
\label{sec:intro}

Traditional embedded systems comprise of a micro-controller, sensors, and actuators. These are tightly coupled in a chip and communication between these modules happens using field bus. Cyber-Physical System (CPS) is an extension of embedded systems, wherein these components connected together in an environment and automatically exchange the information through high speed networks to achieve a task in an application \cite{Ahmed2013CyberChallenges}. To set a tone, the authors considered Smart Industries i.e ICS is an application of CPS in the context of discussing the network layer aspects for CPS security. Industry 4.0 defines the growing trend towards automation and data exchange in technology and processes within the manufacturing industry, including Internet of Things (IoT), CPSs, smart manufacture, cloud computing, and artificial intelligence \cite{Lee2015ASystems}. The future smart industries handle maintenance, production itself and predict the benefit/loss, and optimize resource, energy etc. To meet the requirements of Industry 4.0, a fast, reliable, and secured network is also mandatory to achieve the automation and data exchange. 

Software Defined Networking (SDN) came into limelight for managing the data centers with centralized view of the network infrastructure \cite{Molina2018Software-definedSurvey}. Owing to the advantages that SDN offers such as security, programmability, the researchers also explored to use of SDN in other areas like IoT,  CPS, and enterprise network management in cloud \cite{Grigoryan2018} \cite{Sdwan}. In particular, the network security solutions can be implemented to handle the Denial of Service (DoS), route hijacking , topology poisoning attacks using SDN \cite{Zhu2018} \cite{Hong_poisoningnetwork} \cite{Dhawan2015}. However, an adversary can also leverage the centralized nature of the architecture to saturate the controller and dataplane communication channel or openflow switch flow table flooding or controller resource consumption and perform successful DoS attacks. Some of the solutions proposed to detect and mitigate the denial of service attempts in SDN \cite{Boppana2020}\cite{raj2016utsa}.In CPS applications, the attack vector will be broader\cite{Gao2014AnalysisSystems} and have more ways an adversary can compromise the application infrastructure.  Hence, in complex network environment with multi-network protocol traffic like CPS, there is a need to control how the packets should be processed and controlled in the network layer and improve the monitoring, security capability.  The main contributions of this paper is as follows 
\begin{itemize}
    \item Present an Intelligent network layer architecture using programmable dataplanes to configure, manage and improve the overall Cyber Physical Systems security posture by considering ICS application.
    \item Discuss Hello flooding network attacks detection workflow using P4 programming language as a case study with minimal effort in network layer.
    \item Highlight the importance of advancements in network layer and potential to adopt the programmable dataplanes for Cyber Physical System applications in future.
\end{itemize}
The remainder of the paper is as follows. Section \ref{sec:related} explains the paper background and related work in accordance with our contributions in the paper. Section \ref{sec:cps} discusses a typical Cyber-Physical System architecture. Section \ref{sec:dataplane} illustrates the proposed programmable network device based architecture for CPS security. Section \ref{sec:workflow} highlights the usage of P4 programming to detect HELLO Flooding attack at the network layer. Section \ref{sec:conclusion} concludes the paper and discuss the future work.

\section{Background and Related Work}
\label{sec:related}
In this section, the authors discusses the state-of-the art papers in CPSs focused on security; general discussion on the sensor node networks describing the HELLO Flood Attacks; introduction to Software defined networks and the related work of using SDN architecture in the context of CPS applications.
The most notable malware worm “Stuxnet” was first identified in 2010 targeting Industrial Control Systems(ICS); the malware 
sophistication demonstrated that the physical access can have catastrophic impact on the communication, computation and physical process-based system \cite{Kimzetter2014AnWIRED}. So, the additional physical attributes in CPS poses new security threats in CPS applications like Industrial Control Systems(ICS), Autonomous Vehicular Technologies , Smart Grid, Unmanned Aerial Vehicles (UAV) and Smart Health \cite{Molina2018Software-definedSurvey} and a successful attack exploitation  can have significant social, economic and political impact on the society. The tightly bounded nature of the computational and the physical elements in CPS makes challenging to combat attacks. The security concerns of CPSs got an attention in academia few years ago and several works highlighted the security aspects to secure the CPS applications \cite{Gawanmeh2018TaxonomyApplications}\cite{Bhamare2020CybersecuritySurvey}\cite{Gao2014AnalysisSystems}\cite{Lu2015AFields}
\cite{Nazarenko2019SurveySystems}\cite{Wurm2017IntroductionPerspective}\cite{Darabseh2019ASystems}. In \cite{Gawanmeh2018TaxonomyApplications}, the authors classified the state of the art security solutions for CPS based on the security analysis type (modeling, detection, prevention, response) and  their application. Yang Gao et al. \cite{Gao2014AnalysisSystems} described the security threats and attacks in CPSs by categorizing the threats in accordance with CPS architecture layers such as physical, network and application layer. They also classified the vulnerabilities in CPS as per the management and policy, platform and network. Tianbo Lu and Artem emphasized the security and privacy issues in CPSs and also covered the existing defense mechanisms to defend the security attacks \cite{Lu2015AFields}\cite{Nazarenko2019SurveySystems}. Deval et al. focused on highlighting the security issues in Industrial Control Systems particularly shifting the ICS to cloud environment and discussed the applicability of machine learning based security solutions in such scenarios \cite{Bhamare2020CybersecuritySurvey}. In \cite{Wurm2017IntroductionPerspective}, the authors investigated the security vulnerabilities in the deployed smart home device Nest Thermostat at various layers like device, hardware and countermeasures has been recommended to smart home device manufacturers.

Wireless Sensor Networks(WSN) comprises the sensors, actuators or controllers to communicate among the network nodes and perform the desire action. The participating nodes usually less power operated,memory constrained with communication module; these nodes are prone to network attacks and shows little resilience to defend the attacks by itself. The most notable and well known attacks in wireless sensor networks are Hello Flooding, Sinkhole and Wormhole attacks\cite{Karlof2003SecureCountermeasures}. An attacker may compromise the legitimate nodes and control the other nodes communication for malicious purpose including listen the legitimate node messages, hijacking the traffic routes, manipulating the node messages etc. Hello flooding attack is considered for the scope of this work and other attacks are not detailed in the paper.

In Hello Flooding attack, The legitimate nodes send HELLO messages to malicious nodes believing that they are neighbour nodes, even though the malicious nodes are operating far away from the legitimate nodes. Later, the legitimate nodes pass the messages to malicious and results in packet loss or unauthorized data access. Some of the works addressed wireless sensor networks attacks are implementing the intrusion detection system monitoring framework to capture the network traffic and apply the machine learning algorithms to determine the attack \cite{Napiah2018CompressionProtocol}, applying location verification algorithms to detect the malicious node locations \cite{Hassoubah2015IntrusionScheme}, measuring the signal strength of the nodes as well packet ping response delay times to determine if the node is located nearest to the legitimate node or not \cite{Singh2013SignalNetworks} and the implementation of secured routing protocols based on cryptography mechanisms to verify the node identity. However, these techniques require memory and power requirements, which are critical in wireless sensor network nodes. So, there is a requirement of handling network attacks with no computational and memory burden on the end devices..

In general, traditional networks are not well suited for dynamic, heterogeneous and tightly coupled with physical device network environments. A delayed network connection between communication components and the physical devices like actuators will have a significance impact in CPS application like Industrial Control Systems (ICS) \cite{Bhamare2020CybersecuritySurvey}. Another requirements is that the global view of the infrastructure to control, monitor flow of the information in network stack across the architecture layers. Software Defined Networks(SDN) is a network paradigm, which decouples the control plane from the data plane. SDN offers advantages such as programmability, efficient network management, network global visibility and security \cite{Molina2018Software-definedSurvey}. The open source network protocol Openflow acts as a communication channel between the control plane and data plane and carry the control plane instructions to update the flow tables in data plane network devices.  Elias Molina et al. surveyed the state of the art Software Defined Networks approaches applied to CPS applications, challenges and opportunities \cite{Molina2018Software-definedSurvey}. Ala’ Darabseh1 et al. proposed a Software Defined CPS multi layered architecture, in which the control layer is distributed. The authors divided the cyber physical space into several zones. Each zone is controlled by a local controller, the local controller can communicate to the adjacent controller and all the local controllers are controlled by a global controller\cite{Darabseh2019ASystems}. However, these network architectures proposed to use in CPS application utilize the openflow protocol for southbound API; openflow has limited predefined packet fields that can parse and process at the switch level.

\section{Cyber Physical Systems}
\label{sec:cps}
\begin{figure}[t]
\centering
\includegraphics[width=.5\textwidth, height = .25\textheight]{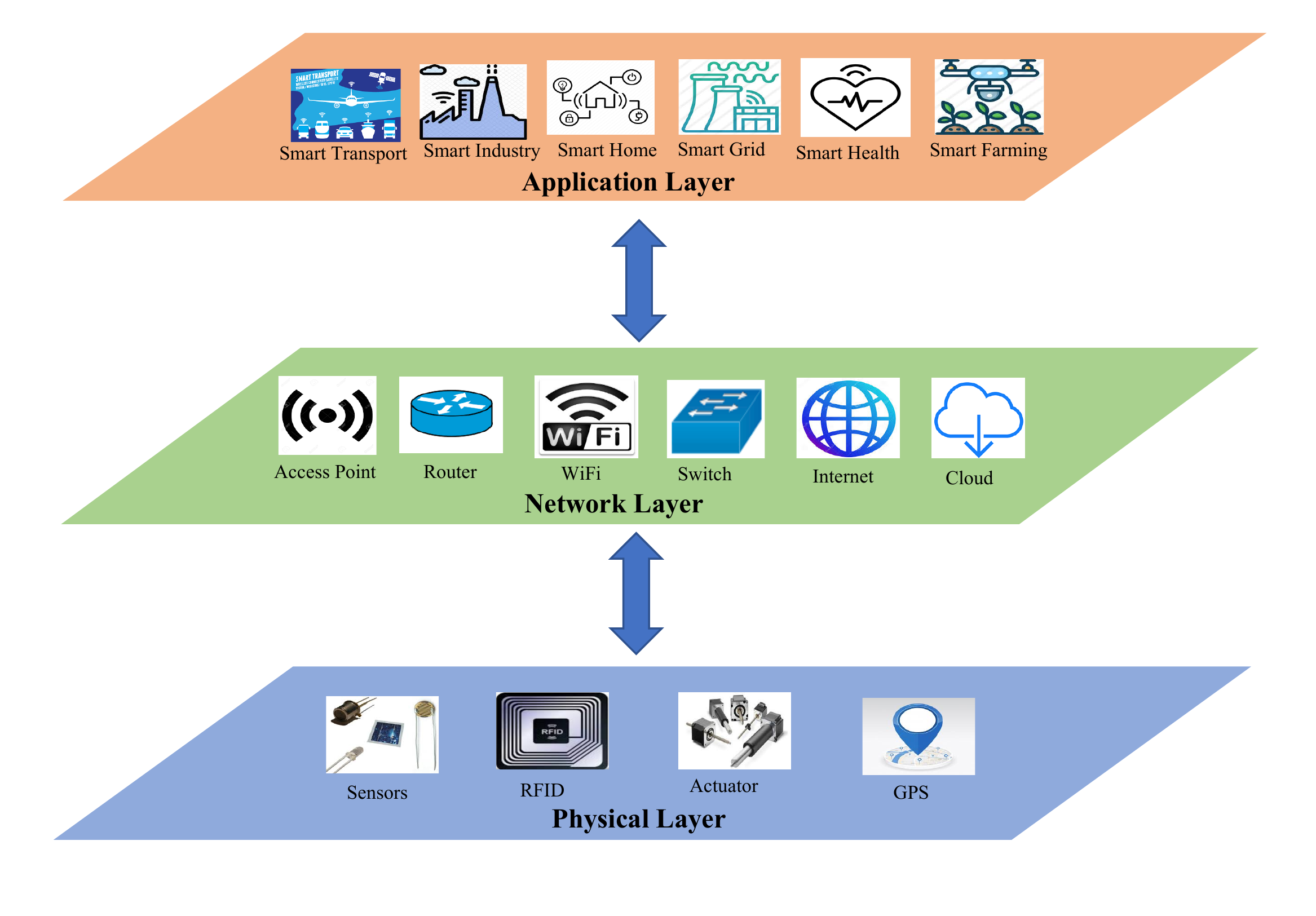}
\centering
\caption{A Typical Layer-Based CPS Architecture.}
\label{fig:cps}
\end{figure}

CPSs consist of the integration of computational systems,
networking and physical processes. This typical architecture
is mainly represented by three main layers, including the
physical, network and application layers, as shown in Figure
1. This layered architecture extends and adapts existing IoT and CPS architectures \cite{gupta2020future, gupta2020security}. 

\subsection{Physical Layer}
The bottom layer of this architecture is a rich set of IoT
devices including sensors, actuators, embedded devices, road
side infrastructures, vehicles, etc. These physical objects are
spread across and implemented in smart communities like
hospitals, retail-stores, homes, parking lots. A smart device
user can directly interact with this layer while using and
controlling the physical objects. Usually, these smart devices
are constrained with limited power, memory, compute, and storage and also deployed in large and unattended environments.

\subsection{Network Layer}
Network communication layer is responsible for establishing the communication between physical sensors, smart devices, edge compute nodes or cloudlets, cloud services and specific application components. This layer usually comprise wireless and wired technologies to get all layers connected and exchange information in CPS. The network layer has multiple threats such as routing attacks, flooding attacks, spoofing attack targeting the systems \cite{Unal2019}. The effective detection and mitigation of these attacks in CPS requires new strategy and methodologies.

\subsection{Application Layer}
This layer delivers specific services to end users through
different CPS applications and communicate through cloud \cite{gupta2020access}. It defines various applications in
which sensors/actuators can be deployed, for example, smart
transport, smart industry, smart homes, and smart grid, etc.
The application layer is more responsible for decision-making
and also activating commands for controlling the physical devices. The application layer attacks such as cross site scripting, SQL Injection, zero day vulnerabilities target to compromise the core applications specific to industry and results in data breaches.

\section{Programmable Data plane Architecture for CPS Security}
\label{sec:dataplane}

\begin{figure*}[t]
\centering
\includegraphics[width=0.70\textwidth, height = .48\textheight]{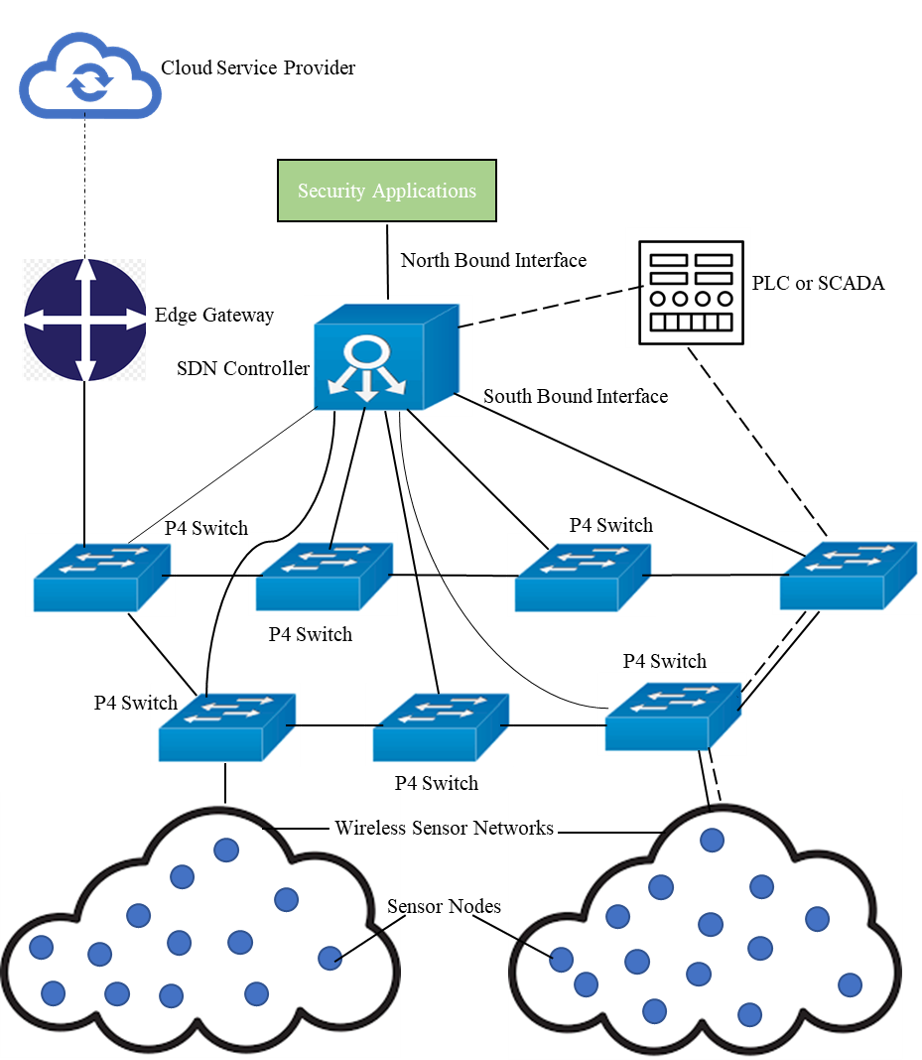}
\caption{Proposed Programmable Data Plane for CPS Security.}
\label{fig:cpsarch}
\end{figure*}
CPSs inherently require security, low latency, and reliable data communication to achieve the desired outcome perhaps controlling the actuator. In addition, the CPS numerous applications possess various network standard protocols for data transfer and management of the network devices is difficult for the network administrators. For example, the wired and wireless communication protocols such as Ethernet, 6Lowpan, Zigbee, Bluetooth are widely used in Industrial control systems to record the sensor data, efficient time critical data transfer between the actuators and controller. To alleviate these problems in the network layer of the CPS, we envision a Programmable dataplanes network layer CPS architecture, which can be used in any of the CPS applications. Programmable protocol independent packet processors(P4) is a programmable language to tell the  dataplanes  how to process the packets irrespective of the network protocols and underlying hardware such as ASIC, FPGA or CPU processing the network packets \cite{BosshartP4:Processors}. 

The figure \ref{fig:cpsarch}  shows the proposed programmable network layer architecture for the CPS application Industrial Control System. The sensor nodes are distributed across the industrial systems to monitor the physical and environmental conditions such as temperature, pressure or measuring the physical and chemicals properties of materials etc; sensed data is aggregated at the base station(Aggregation sensor node), as the nodes can pass the information to neighbor nodes and so on until the destined base station receive the data. These base stations can be gateways to transmit the information. The base station is connected to the P4 switch to process the sensor data packets as per the defined P4 program compiled into the P4 switches. It is also possible to have edge gateway connected to P4 switch for passing the information. In general, the physical systems and operation technology(OT) centers are connected via traditional networks possibly high-speed networks. However, the network administrator has limited control over the switches due to unable to reconfigure the switch program and not able to process new packet headers. Whereas, the P4 switches can forward the packets of interest to controller for security monitoring and flag the users when needed as well as can reprogram the switches to handle multitude of network protocols involved in  CPSs. 
 
SDN controller can communicate P4 switches through southbound API like Openflow to collect the packet specific information or packet stats. The security applications are implemented on top of the controller using machine learning algorithms or statistical algorithms to detect the attacks in CPSs. Applications could be Intrusion detection system, firewall, secured storage to protect the network from cyber attempts. Another advantage of using centralized network monitoring with p4switches is to correlate the security events in operation technology(OT) infrastructure environment with the ICS information technology(IT) events. The adversaries usually  choose the easiest path to compromise the ICS  IT infrastructure by phishing the employees and then perform lateral movement to infect the PLC or SCADA infrastructure in ICS. As the workforce managed the security of these platforms in ICS are not the same, it is very likely to evade the security detection mechanism and impact the physical systems as well. We can envision the future P4 based ICS architectures with East and West bound API’s sharing  the IT and OT environment data and securing the ICS with unified view of the network data flow.  

The PLC or SCADA components are connected to the sensor or actuators nodes through P4 switches, which can handle line rate data packet processing and suitable for delay sensitive CPS applications. As shown in figure 2, the dotted lines shows a direct connection from SCADA systems to sensor nodes and the SDN controller. There are numerous possibilities to build a data analytics platform on top of SDN controller. For instance, the north bound interface API calls to Amazon web services S3 storage to load the data and leverage the big data analytics platforms  in cloud. In addition, the vendor specific devices like sensors can be connected to the vendor cloud through the programmable dataplane switches to constantly update and manage the devices online. In this way, our advanced network layer architecture would benefit the heterogeneous and complex environments network management. The advanced connectivity capabilities to real time connection, data analytics and computation capability of this architecture would be added advantage to current state of the art for advancement.

\section{P4 Workflow for Hello Flood Attack Detection}
\label{sec:workflow}

\begin{figure*}[t]
\centering
\includegraphics[width=.6\textwidth, height = .45\textheight]{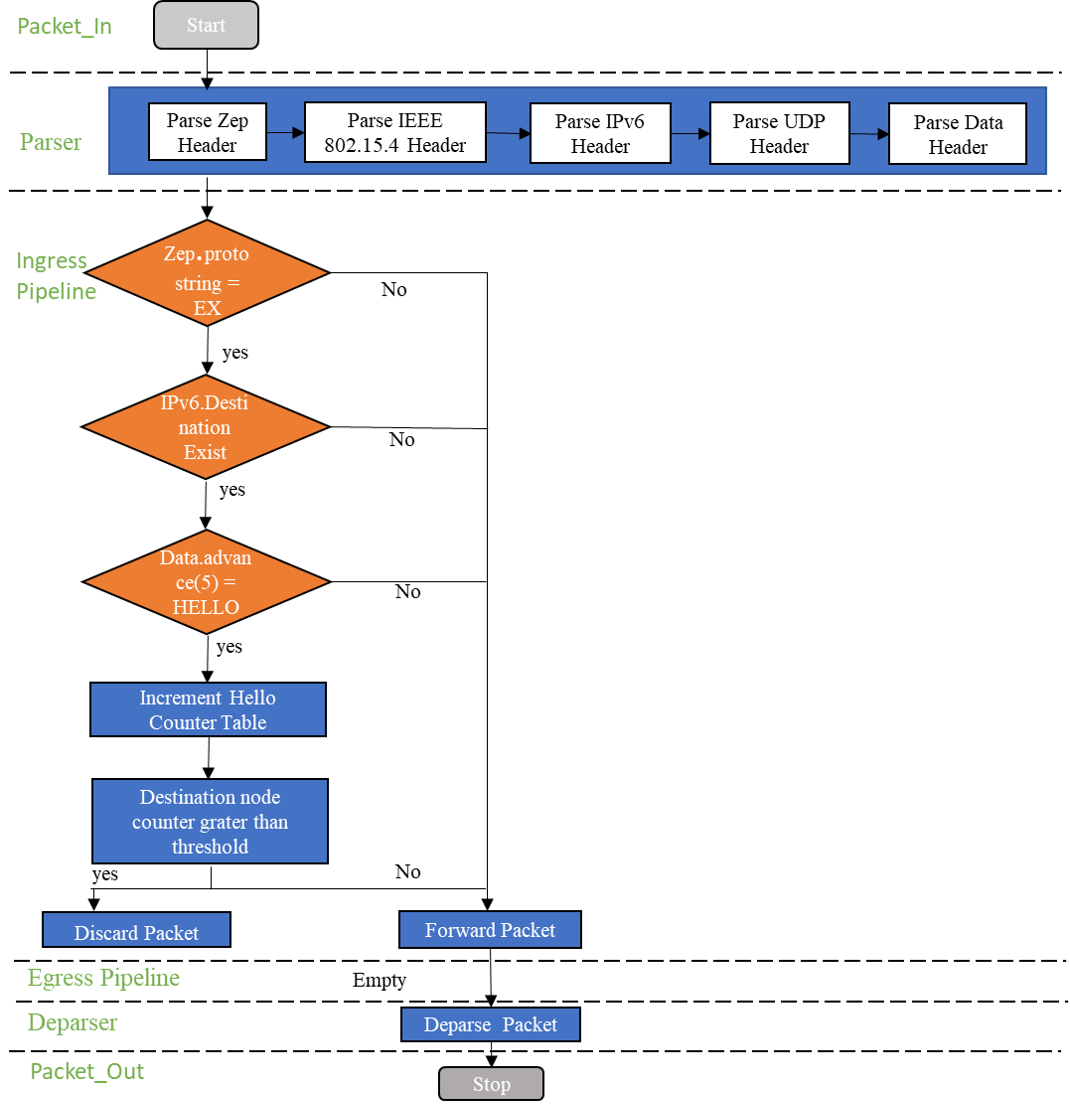}
\centering
\caption{P4 Workflow for HELLO FLOOD Attack Mitigation.}
\label{fig:p4attack}
\end{figure*} 

In this section, we define a workflow in programmable dataplanes switches to detect the malicious nodes performing HELLO Flood attacks in the ad-hoc sensor networks, which are part of CPS; assess the potential capability of defending the network security attacks in CPS by leveraging the programmable Dataplane network architecture. 

The network domain specific language P4 is a formal high-level description of how the packet headers are processed and the actions to be performed on the extracted packet header fields. We have leveraged P416 version to describe the P4 usage in the paper. P4 program consists of four main components. Header describes the sequence and structure of a series of header fields. We have considered  Zigbee(Zep), IEEE 802.15.4, 6LOWPAN, UDP  and Data headers to extract the 6LOWPAN network packets in P4 switch. For example, The IEEE 802.15.4 is described as follows and being used to parse the IEEE 802.15.4 packets.\\ \newline
Header IEEE802154 \{ \\
\hspace*{10mm} framecontrol: 16 \\
\hspace*{10mm} Seqnumber: 8 \\
\hspace*{10mm} DestPAN: 16 \\
\hspace*{10mm} Destination: 48 \\
\hspace*{10mm} ExtendSrc: 48 \\
\hspace*{10mm} FCS: 16 \\
\} \newline
The second component Parser is used to identify the headers and header fields in the incoming packet to the P4 switch. When the packet received at the Ingress port in P4 switch, the parser starts in the “start” state and then parses Zigbee packer header. If the Zigbee packer field “ProtoIDString” is equal to “EX”, the parser goes to the next header IEEE802154 parsing and so on until the end state. \newline \\
\lstset{numbers=left, numberstyle=\tiny, stepnumber=1, numbersep=5pt}
State Start \{ \\
\hspace*{10mm} transition Zigbee; \\ 
\} \\ 
State Zigbee \{ \\
\hspace*{10mm} packet.extract(hdr.zep)    \\  
\hspace*{10mm} transition select(hdr.zep.ProtoIDstring)\{ \\
	\hspace*{20mm} 0x4548: IEEE802154; \\
 \hspace*{10mm} \} \\
\} \newline
When parsing the packet headers, the header fields can be assigned to user defined variable with bit size to perform operations on those header fields before departing and egress out to the switch. 
P4 program also includes the match-action tables and whenever there is a match of specific field in the table, the corresponding defined action is performed on the packet fields. For instance, to perform the network address translation functionality in P4 Switches, the match key is chosen as IP address and the corresponding primitive action function set\_field\(\) to set the IP address in the IP header. We monitor the adversarial nodes receiving HELLO packets and incremented the count of the number of packets being received as shown below match-action table. Whenever a packet is received into the switch, this table matches the destination address of 6Lowpan header. If there is a match, the node corresponding counter will be incremented. The size of table can be defined by the programmer based on the number of nodes connected to the network for tracking.\\ \newline 
table dstnodecounter \{ \\
\hspace*{10mm} actions = \{  \\
\hspace*{15mm}  counterincr; \\
\hspace*{10mm}	\} \\
\hspace*{10mm}	Key = \{ \\
\hspace*{15mm}	hdr.6lowpan.dst: exact;  \\
\hspace*{10mm}	\} \\
 \hspace*{10mm} Size=512; \\
\} 

The third component is control program to define the control flow of the program. The flow of the program control is defined in main function to follow the packet processing pipeline, which is described in the figure 3. For example, the typical pipeline used for BMV2 software switch is as follows.\newline 
\hspace*{10mm} Parser()  \\
\hspace*{10mm} verifychecksum()  \\
\hspace*{10mm} Ingress()  \\
\hspace*{10mm} Egress()  \\
\hspace*{10mm} computerchecksum()  \\
\hspace*{10mm} Deparser()  \newline
The counter table will be executed in Ingress() module of the packet processing pipeline.\\ \newline
Control ingress \{ \\
\hspace*{10mm} Apply \{ \\
\hspace*{20mm}	Dstnodecounter.apply();	\\
\hspace*{10mm}       \} \\
\} 

The checksum module was not used in our workflow, as we don’t need to replace header fields in the packet. \newline
The P4 programs need to be compiled to translate into Table Dependency Graphs(TDG) and then mapping the TDG on to specific hardware to allocate the memory and other resources. The performance and security of the proposed workflow in P4 switch is evaluated using Mininet emulator \cite{mininet} and BMV2 \cite{bmv2} P4 software switch as the next steps of the workflow.

\section{Conclusion and Future work}
\label{sec:conclusion}
In the past, it is very rare to see the network architectural changes once build the application, as the technological modifications require build the network from scratch, high cost, and extensive workforce. The technical requirements  in CPSs, IoT and Machine Learning demands sophisticated network infrastructure to deal with security, latency, computation, and data processing capabilities. This paper proposed programmable network dataplane based network architecture for implementing CPS applications and in specific focusing on ICS. To the end, a security application HELLO Flood attack detection is implemented using P4 programming procedure and semantics as a case study to showcase the programmable dataplanes are good candidate for CPS network layer. Our future work includes evaluating the performance overhead of the proposed workflow to detect and mitigate the flooding attacks using Mininet emulation environment and define a framework to detect all typical network attacks in CPS applications with the programmable network layer.


{
\bibliographystyle{unsrt}
\bibliography{references}
}
\end{document}